\documentclass[conference]{IEEEtran}

\usepackage{slashbox}
\usepackage{cite}
\usepackage{algorithmicx}
\usepackage{color}
\usepackage{algorithm}
\usepackage{algpseudocode}
\usepackage{amssymb}
\usepackage{authblk}
\ifCLASSINFOpdf
   \usepackage[pdftex]{graphicx}

\fi

\usepackage{amsmath,bm}
\usepackage{setspace}
\usepackage{array}

\IEEEoverridecommandlockouts
\begin{document}
\title{Bayesian Estimation Based Parameter Estimation for Composite Load}
\author[1, 2]{Chang Fu}
\author[1]{Zhe Yu}
\author[1]{Di Shi}
\author[3]{Haifeng Li}
\author[2]{Caisheng Wang}
\author[1]{Zhiwei Wang}
\author[4]{Jie Li\thanks{This work is funded by SGCC Science and Technology Program under contract no. SGSDYT00FCJS1700676.}}

\affil[1]{GEIRI North America, San Jose, CA, 95134, USA}
\affil[2]{ College of Engineering, Wayne State University, Detroit, MI, 48201, USA}
\affil[3]{State Grid Jiangsu Electric Power Company Ltd., Nanjing, Jiangsu, 210024, China}
\affil[4]{State Grid US Representative Office, New York City, NY, 10017, USA  \authorcr Email: {  zhe.yu@geirina.net}}
\maketitle

\begin{abstract}
Accurate identification of parameters of load models is essential in power system computations, including simulation, prediction, and stability and reliability analysis. Conventional point estimation based composite load modeling approaches suffer from disturbances and noises and provide limited information of the system dynamics. In this work, a statistic (Bayesian Estimation) based distribution estimation  approach is proposed for both static (ZIP) and dynamic (Induction Motor) load modeling. When dealing with multiple parameters, Gibbs sampling method is employed. In each iteration, the proposal samples each parameter while keeps others fixed. The proposed method provides a distribution estimation of load models coefficients and is robust to measurement errors.
\end{abstract}

%
\IEEEpeerreviewmaketitle

\section{Introduction}
\label{introduction}
\IEEEPARstart{L}{oad} models can be categorized into static and dynamic models. Conventional static models include ZIP model, exponential model, frequency dependent model, \emph{etc.}, and the induction motor (IM), exponential recovery load model (ERL) \cite{08} are the major dynamic load models used in research studies recently. It is reported that a composite load under ZIP+IM model can be used with various conditions, locations and composition, and it has been widely used in industry and studied in voltage stability as well as planning, operation and control of power systems\cite{08,book2,11,32}. \par


 Load models are typically identified by component and measurement based approaches in state of the  art \cite{08,09,gibbsmjin, expzip1,10,16,17,33,desong,YuEtalTSG2012,30}.  
Component based approaches highly rely on characteristics of components and compositions of  load. In \cite{expzip1}, ZIP coefficients for widely used electrical appliances are determined by experiments, and the overall ZIP model was established with respect to the predetermined appliances. However, the computational cost is high, and the accuracy of individual consumers is greatly affected by non-electric factors such as data and weather \cite{expzip1,08}. Difficulties in obtaining  load composition information is another factor  to be considered when implementing this type of methods.\par
  Measurement-based approaches, such as least-squares (LS) and genetic algorithm (GA), are another mainstream for load parameter identifications.
Different techniques are reported in \cite{16,21} to identify the parameters for composite load models. In \cite{15}, a robust time-varying load parameters identification approach is proposed by using  batch-model regression in order to obtain the updated system parameters.
One disadvantage of measurement based approach is the dependence on data quality. Measurement anomalies may affect the robustness of estimation in both time-varying and time-independent load models. Moreover, these approaches estimate the expected value of coefficients, which provides  limited information.

Bayesian estimation (BE) \cite{29} based composite load parameter identification approach can successfully overcome the aforementioned disadvantages in measurement and component based approaches. First, BE is a distribution estimation rather than point estimation technique,
 which provides the likelihood of  each parameter. In this case, this method is able to provide accurate estimation of the parameters in both time constant and time-varying cases. Second, BE does not require information of load compositions, nor coefficients of appliances. Third, BE is a robust estimation method due to its statistical characteristics: measurement anomalies will not significantly affect the results if the number of samples is large enough, and the effect of  measurement error is also not significant either because the distribution interval contains the real value inside the distribution with expected probabilities. \par

The rest of the paper is organized as follows. The problem of ZIP and IM model estimation is formulated in Section \ref{sec:formulation}. The Bayesian estimation technology and the detailed parameter identification method (Gibbs Sampling) is introduced in Section \ref{sec:BE}. The advantage of the proposal is illustrated in Section \ref{casestudy} by simulations. Section \ref{conclusion} draws the conclusion.
\section{Formulation}
\label{sec:formulation}
In this section, a composite load model consisting of ZIP and IM model is introduced \cite{book2}, followed by the proposed parameter estimation method, Gibbs Sampling (GS).

\begin{figure}[tbh]
\centering
    \includegraphics[width=0.7\linewidth,]{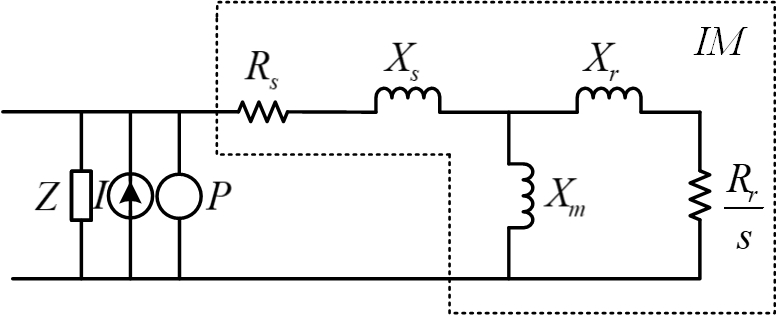}
\caption{The composite model of ZIP+IM.}
\label{composite}
\end{figure}

A composite load with ZIP+IM model is shown in Fig. \ref{composite}, in which the ZIP model describes the steady-state behavior and the IM model corresponds to the dynamic process.
\subsection {ZIP Model}
The ZIP model describes how the  power of load changes as voltage varies in the steady-state condition. The  ZIP model is formulated as follows.
\[
\begin{array}{l}
P=P_0(\alpha_1\bar{V}^2+\alpha_2\bar{V}+\alpha_3), Q=Q_0(\alpha_4\bar{V}^2+\alpha_5\bar{V}+\alpha_6),
\label{zipformula}
\end{array}
\]
%
%
%
%
where ${\sum_{i=1}^3{\alpha_i}=\sum_{i=4}^6{\alpha_i}=1}$, $\bar{V}=V/V_0$, $P$ and $Q$ are real and reactive power, and $V$ is voltage magnitude at the load terminal. Variables $P_0$, $Q_0$ and $V_0$ represent the values of the respective variables at the initial operating condition. When voltage $V$ deviates from $V_0$, real and reactive power of the load is assumed to follow a quadratic model. As the real power and reactive power follow the same form of model,  we will discuss only the real power as an example in the following. The reactive power follows the same procedure.

Define ${y[t]=P[t]/P_0}$, ${x[t]=V[t]/V_0}$, where $P[t]$ and $V[t]$ are the $t$th measurements of the real power and voltage magnitude of the ZIP load. The following assumptions are made.
\begin{itemize}
\item The measurement noise follows a normal distribution, \emph{i.e.}, ${y[t]=\alpha_1x[t]^2+\alpha_2x[t]+\alpha_3+\varepsilon[t]}$,
where ${\varepsilon[t]\sim \mathcal{N}(0,1/\tau)}$, and $1/\tau$ is the variance\footnote {The reasons for making this assumption are: 1) According to the law of large numbers, a normal distribution would be the best one to represent the characteristics of the noise if the number of experiment is large enough. 2) Since normal distribution is a conjugate distribution, it is easier for model parameter updating when implementing Gibbs sampling.}.
\item Total number of $n$ independent and identically distributed (i.i.d.) samples were drawn, namely ${(\textbf{x},\textbf{y})\triangleq\{(x[1],y[1]),\cdots,(x[n],y[n])\}}$. Thus the likelihood
\begin{equation}\label{likelihood}
p(\textbf{x},\textbf{y}|\alpha_1,\alpha_2,\tau)\propto\prod_{t=1}^n\exp[(y[t]-\mu[t])^2\tau/2],
\end{equation}
where $\mu[t]\triangleq\alpha_1x[t]^2+\alpha_2x[t]+1-\alpha_1-\alpha_2$ is the mean.
\end{itemize}

 \subsection{ IM Model} The dynamic part of load is usually represented by an IM  model, which is discussed below \cite{gibbsmjin}.
\[
\begin{array}{l}
\left\{
\begin{array}{l}
dE'_d/dt=-[E'_d+(X-X')I_q]/T'-(\omega-1)E'_q\\
dE'_q/dt=-[E'_q-(X-X')I_d]/T'+(\omega-1)E'_d\\
d\omega/dt=-[(A^2\omega+B\omega+C)T_0-(E'_dI_d+E'_qI_q)]/2H
\end{array}
\right.\\
\left\{
\begin{array}{l}
I_d=[R_s(U_d-E'_d)+X'(U_q-E'_q)]/(R_s^2+X'^2)\\
I_q=[R_s(U_q-E'_q)-X'(U_d-E'_d)]/(R_s^2+X'^2)
\end{array}
\right.
\end{array}\]
where ${X'\triangleq X_s+X_mX_r/(X_m+X_r)}$, ${X\triangleq X_s+X_m}$,  ${T'\triangleq(X_r+X_m)/R_r}$, ${A+B+C=1}$.
Here $R_s$ is motor stator winding resistance, $X_s$ is motor stator leakage reactance, $X_m$ is motor magnetizing reactance, $R_r$ is  rotor resistance, $X_r$  is  rotor leakage reactance, $H$ is  rotor inertia constant, $\omega $ is  rotor speed, $I_d$ and $I_q$ are stator current in $d$-axis and $q$-axis, $U_d$ and $U_q$ are  bus voltage in $d$-axis and $q$-axis, and $E'_d$ and $E'_q$ are stator voltage in $d$-axis and $q$-axis. $T_0$ is the initial load torque.

The parameters to be identified in an IM model are: $X_r, X_m, X_s, R_r, R_s, A, B, C, H$. Typically, $A$ is assumed to be $1$ as the mechanical torque is assumed to be proportional to the square of the rotation speed of the motor \cite{17}. Consequently B and C are both equal to 0. For simplicity,  define $y_{Ed} \triangleq dE'_q/dt$, $y_{Eq} \triangleq dE'_d/dt$, $y_{\omega}\triangleq d\omega/dt$, $ y_{Id}\triangleq I_d$, $ y_{Iq}\triangleq I_q$, $ \beta_1\triangleq-1/T'$, $\beta_2\triangleq-(X-X')/T'$, $\beta_3\triangleq-{1}/{2H}$, $\alpha_b \triangleq {R_s}/{R_s^2+X'^2}$, $\alpha_c \triangleq{X'}/{R_s^2+X'^2}$. Assuming that the measurement noise is i.i.d. and follows a normal distribution, we can rewrite the IM model as follows:
\begin{equation}\label{equa1}\begin{array}{l}
\left\{
\begin{array}{l}
y_{E_d}[t]=\beta_1E'_d[t]+\beta_2I_q[t]-(\omega-1)E'_q[t]+\varepsilon_{E_d}[t]\\
y_{E_q}[t]=\beta_1E'_q[t]-\beta_2I_d[t]+(\omega-1)E'_d[t]+\varepsilon_{E_q}[t]\\
y_{\omega}[t]=\beta_3(\omega^2-E'_d[t]I_d[t]-E'_q[t]I_q[t])+\varepsilon_{\omega}[t]\\
\end{array}
\right.\\
\left\{
\begin{array}{l}
y_{I_d}[t]=\alpha_b(U_d[t]-E'_d[t])+\alpha_c(U_q[t]-E'_q[t])+\varepsilon_{I_d}[t]\\
y_{I_q}[t]=\alpha_b(U_q[t]-E'_q[t])+\alpha_c(U_d[t]-E'_d[t])+\varepsilon_{I_q}[t]
\end{array}
\right.
\end{array}\end{equation}
where ${\varepsilon_{E_d}[t]\sim \mathcal{N}(0,1/\tau_{E})}$, ${\varepsilon_{E_q}[t]\sim \mathcal{N}(0,1/\tau_{E})}$, ${\varepsilon_{\omega}[t]\sim \mathcal{N}(0,1/\tau_{\omega})}$, ${\varepsilon_{I_d}[t]\sim \mathcal{N}(0,1/\tau_{I})}$, and ${\varepsilon_{I_q}[t]\sim \mathcal{N}(0,1/\tau_{I})}$.

\section{Bayesian Estimation in Composite Load Parameter Identification}\label{sec:BE}
\subsection{Gibbs sampling}
Gibbs sampling is an extension of Monte Carlo Markov Chain method \cite{29}, which performs well when there are multiple parameters to identify. The detailed sampling algorithm is shown in Algorithm \ref{gibbs}.
\makeatletter
\def\algbackskip{\hskip-\ALG@thistlm}
\makeatother

\begin{algorithm}[!h]
\caption{Gibbs Sampling}\label{gibbs}
\begin{algorithmic}[1]
\State Draw initial samples $\bm{\theta}^{(0)} \sim {q(\bm{\theta})}$, where $q(\bm{\theta})$ is the prior.
\For {$\text{iteration}\ i=1,2,...,M$}
\State Calculate $p(\theta_1|\theta_2^{(i-1)},\theta_3^{(i-1)},...,\theta_n^{(i-1)})$ and sample $\theta_1^{(i)}\sim p(\theta_1|\theta_2^{(i-1)},\theta_3^{(i-1)},...,\theta_n^{(i-1)})$


 $\vdots$
\State Calculate $p(\theta_n|\theta_1^{(i)},\theta_3^{(i)},...,\theta_{n-1}^{(i)})$ and sample ${\theta_n^{(i)}\sim p(\theta_n|\theta_1^{(i)},\theta_3^{(i)},...,\theta_{n-1}^{(i)})}$
\EndFor
\State  The distribution estimate is the histogram of $\bm{\theta}^i$, ${i=m,\cdots,M}$. Others are burn-in data and discarded.
\end{algorithmic}
\end{algorithm}
 Starting with priors, Gibbs sampling estimates the posterior of one parameter while fixing others' values as samples from previous estimated posteriors. This process repeats for all parameters in one iteration.

\subsection{Gibbs in ZIP Model}\label{gibbsinzipmodel}

Start with the prior guess as follows:
\begin{eqnarray}\label{mu1}
\alpha_1&\sim& \mathcal{N}(\mu_{1}^{(0)},1/\tau_{1}^{(0)}),\\
\alpha_2&\sim& \mathcal{N}(\mu_{2}^{(0)},1/\tau_{2}^{(0)}),\\
\tau&\sim& \mathcal{G}(a^{(0)},b^{(0)}).
\end{eqnarray}
where the distribution of $\tau$ is a gamma distribution follows $\mathcal{G}(a, b)$. It can be shown that after each iteration of Gibbs sampling, the post distributions of these three parameters remains the same form. Only the first iteration will be shown here as an example.
First,  samples are drawn from the prior and $\alpha_{1}^{(0)},\alpha_2^{(0)},\tau^{(0)}$ are initialized. After some algebraic yields:
\begin{equation}
\label{posterior}
p(\alpha_1|\alpha_2^{(0)},\tau^{(0)},\textbf{x},\textbf{y})\propto p(\textbf{x},\textbf{y}|\alpha_1,\alpha_2^{(0)},\tau^{(0)})p(\alpha_1)
\end{equation}
where $p(\alpha_1|\alpha_2^{(0)},\tau^{(0)},\textbf{x},\textbf{y})$ is the posterior probability given the samples of $\textbf{x}$, $\textbf{y}$, $\alpha_2$, and $\tau$, $p(\textbf{x},\textbf{y}|\alpha_1,\alpha_2^{(0)},\tau^{(0)})$ is the likelihood in (\ref{likelihood}), and $p(\alpha_1)$ is the prior estimation following (\ref{mu1}).  Taking the log form on both sides of  (\ref{posterior}) yields
\begin{equation}\label{logalpha1}
\begin{array}{l}
\log  p(\alpha_1|\alpha_2^{(0)},\tau^{(0)},\textbf{x},\textbf{y})\propto\\
~~-\dfrac{\tau_{1}^{(0)}}{2}(\alpha_1-\mu_{1}^{(0)})^2-\\
~~\dfrac{\tau^{(0)}}{2}\sum_{t=1}^n\big(y[t]-(\alpha_1x[t]^2+\alpha_2^{(0)}x[t]+1-\alpha_1-\alpha_2^{(0)})\big)^2
\end{array}
\end{equation}

Taking log form helps convert the multiplications of the probabilies to summations, which can significantly simplify calculations when updating the distributions of the posteriors. For a normal distribution $y\sim \mathcal{N}(\mu,1/\tau)$, the log dependence on $y$ is $-\frac{\tau}{2}(y-\mu)^2\propto -\frac{\tau}{2}y^2+\tau \mu y$.

The right hand side of equation (\ref{logalpha1}) can be further written as the following if the terms not related to  $\alpha_1$ are omitted.
\[
\begin{array}{l}
-\big(\tau_{1}^{(0)}+\tau^{(0)}\sum_{t=1}^n(x[t]^2-1)^2\big)\alpha_1^2/2+\big(\tau_1^{(0)}\mu_1^{(0)}\\
~~-\tau^{(0)}\sum_{t=1}^n(\alpha_2^{(0)}-1-\alpha_2^{(0)}x[t]+y[t])(1-x[t]^2)\big)\alpha_1
\end{array}
\]
Define
\[
\hspace{-0.5em}
\begin{array}{l}
\mu_1^{(1)}\triangleq\\
\dfrac{\tau_1^{(0)}\mu_1^{(0)}-\tau^{(0)}\sum_{t=1}^n\big((\alpha_2^{(0)}-1-\alpha_2^{(0)}x[t]+y[t])(1-x[t]^2)\big)}{\tau_1^{(0)}+\tau^{(0)}\sum_{t=1}^n(x[t]^2-1)^2},\\
\end{array}
\]
\[
\begin{array}{l}
\tau_1^{(1)}\triangleq\tau_1^{(0)}+\tau^{(0)}\sum_{t=1}^n(x[t]^2-1)^2,
\end{array}
\]
yields
\[
\alpha_1|\alpha_2^{(0)},\tau^{(0)},\tau_1^{(0)},\mu_1^{(0)},\textbf{x},\textbf{y}\sim \mathcal{N}(\mu_1^{(1)},1/\tau_1^{(1)}).
\]
Then sample a new $\alpha_1$ from the estimated distribution $\mathcal{N}(\mu_1^{(1)},\tau_1^{(1)})$ as $\alpha_1^{(1)}$. Following the similar procedures $\alpha_2$ can be derived. Define
\[
\hspace{-0.5em}
\begin{array}{l}
\mu_2^{(1)}\triangleq\\
\dfrac{\tau_2^{(0)}\mu_2^{(0)}-\tau^{(0)}\sum_{t=1}^n\big((\alpha_1^{(1)}-1-\alpha_1^{(1)}x[t]+y[t])(1-x[t])\big)}{\tau_2^{(0)}+\tau^{(0)}\sum_{t=1}^n(x[t]-1)^2},
\\\tau_2^{(1)}\triangleq\tau_2^{(0)}+\tau^{(0)}\sum_{i=1}^n(x[t]-1)^2.
\end{array}
\]
Therefore, the following can be derived:
\[
\alpha_2|\alpha_1^{(1)},\tau^{(0)},\tau_1^{(0)},\mu_1^{(0)},\textbf{x},\textbf{y}\sim \mathcal{N}(\mu_2^{(1)},1/\tau_2^{(1)}).
\]

%
%

For $\tau$, the posterior given new samples of $\alpha_1^{(1)}$ and $\alpha_2^{(1)}$ can be written as ${p(\tau|\alpha_1^{(1)},\alpha_2^{(1)},\textbf{x},\textbf{y})\propto p(\textbf{x},\textbf{y}|\alpha_1^{(1)},\alpha_2^{(1)},\tau)p(\tau)}$.  Taking the log form of both sides of the posterior and yields
\[
\begin{array}{l}
\log p(\tau|\alpha_1^{(1)},\alpha_2^{(1)},\textbf{x},\textbf{y})\propto\\
~~\dfrac{n}{2}\log \tau-\dfrac{\tau}{2}\sum_{t=1}^n\big(y[t]-\alpha_1^{(1)}x[t]^2-\alpha_2^{(1)}x[t]-\\
~~1+\alpha_1^{(1)}+\alpha_2^{(1)}+(a^{(0)}-1)\log \tau-b^{(0)} \tau\big).
\end{array}
\]
Define
\[
\begin{array}{l}
a^{(1)}=a^{(0)}+n/2,\\
b^{(1)}=b^{(0)}+\\
\sum_{t=1}^n\big(y[t]-\alpha_1^{(1)}x[t]^2-\alpha_2^{(1)}x[t]-(1-\alpha_1^{(1)}-\alpha_2^{(1)})\big)^2/2.
\end{array}
\]
the posterior  $\tau|\alpha_1^{(1)},\alpha_2^{(1)},\textbf{x},\textbf{y}\sim \mathcal{G}(a^{(1)},b^{(1)})$ can be obtained.
%
%
%
%
\subsection{Gibbs in IM models}
Start with the priors as follows.
\begin{eqnarray}
\nonumber
\beta_1&\sim& \mathcal{N}(\mu_{\beta_1}^{(0)},1/\tau_{\beta_1}^{(0)}), \beta_2\sim \mathcal{N}(\mu_{\beta_2}^{(0)},1/\tau_{\beta_2}^{(0)})\\
\nonumber
\beta_3&\sim& \mathcal{N}(\mu_{\beta_3}^{(0)},1/\tau_{\beta_3}^{(0)}), \alpha_b\sim \mathcal{N}(\mu_{\alpha_b}^{(0)},1/\tau_{\alpha_b}^{(0)})\\
\nonumber
\alpha_c&\sim& \mathcal{N}(\mu_{\alpha_c}^{(0)},1/\tau_{\alpha_c}^{(0)}), \tau_{{E}}\sim \mathcal{G}(\alpha_{{E}}^{(0)},\beta_{{E}}^{(0)})\\
\nonumber
\tau_{{\omega}}&\sim& \mathcal{G}(\alpha_{{\omega}}^{(0)},\beta_{{\omega}}^{(0)}), \ \tau_{{I}}\sim \mathcal{G}(\alpha_{{I}}^{(0)},\beta_{{I}}^{(0)}).
\end{eqnarray}
The parameter identification in the IM model follows a similar procedure as the ZIP model and is omitted here for space consideration. Details can be found in \cite{28}.  \par
\section{Case Study}\label{casestudy}
Simulation studies are carried out in this section for ZIP and IM models, separately. In this work, the maximum Monte Carlo runs $M$ is set to 40000 and the burn-in length $m$ is chosen as 5000 \cite{29}.

\subsection{ZIP Model Identification}
  The 33-bus test feeder \cite{34} is used to generate the testing data. A detailed description of the test system can be found in \cite{28}. Load at bus 18 is replaced by a ZIP model. Randomness is added to loads at other buses by multiplying a random factor drawing from a uniform distribution following $\mathcal{U}[0.1,4.5]$. The random load is stated as $P_{w,i}=P_{i}\cdot \mathcal{U}[0.1,4.5]$, where $P_{i}$ is the original load in the 33-bus test feeder, $i$ indicates the node number, and $P_{w,i}$ is the weighted load at the corresponding node after multiplying a random weight.
The changes at each load in every experiment simulates the disturbances and uncertainties in the system, which significantly impact the voltage and power at the bus of interest.

The ZIP factors of the load at node 18 were assigned as ${\alpha_1=0.25}$, ${\alpha_2=0.25}$, ${\alpha_3=0.5}$. Power flow converges in each iteration with different $P_{w,i}$, and the corresponding voltage $V_{w,i}$ is recorded. The measurement noise $\varepsilon$ follows $\mathcal{N}(0,0.1)$, which means a 10\% measurement error.

\begin{figure}[!t]
\centering
    \includegraphics[width=1\linewidth,]{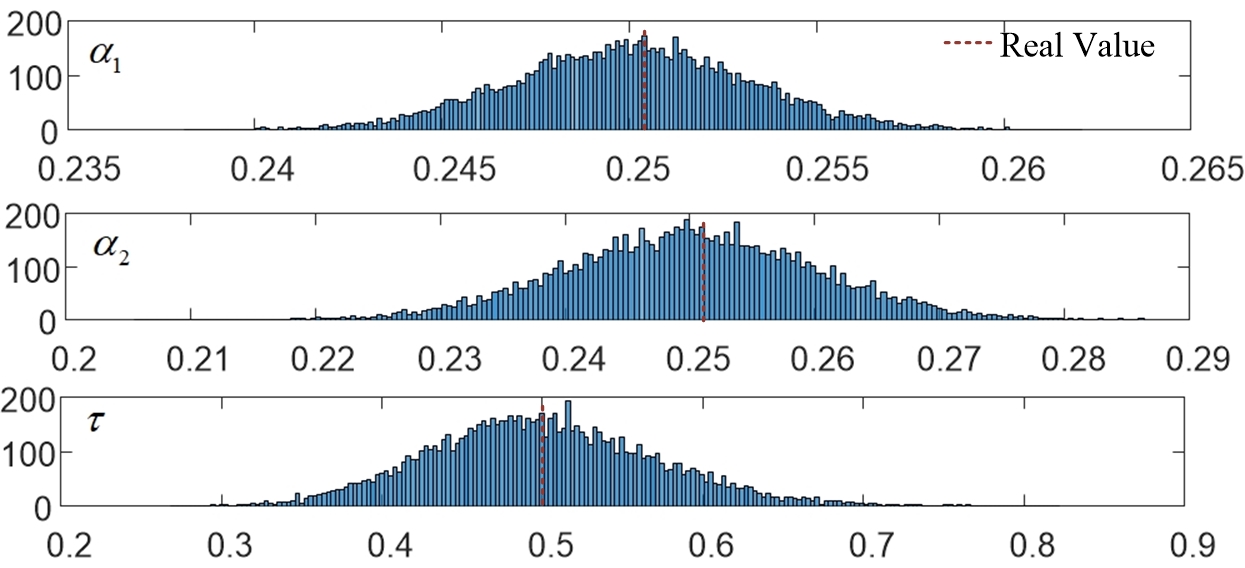}
\caption{The estimated parameters of the ZIP model.}
\label{para1}
\end{figure}

By implementing the proposed GS method, the coefficients of the ZIP model are estimated as shown in Fig. \ref{para1}.
\begin{figure}[!t]
\centering
    \includegraphics[width=1\linewidth,]{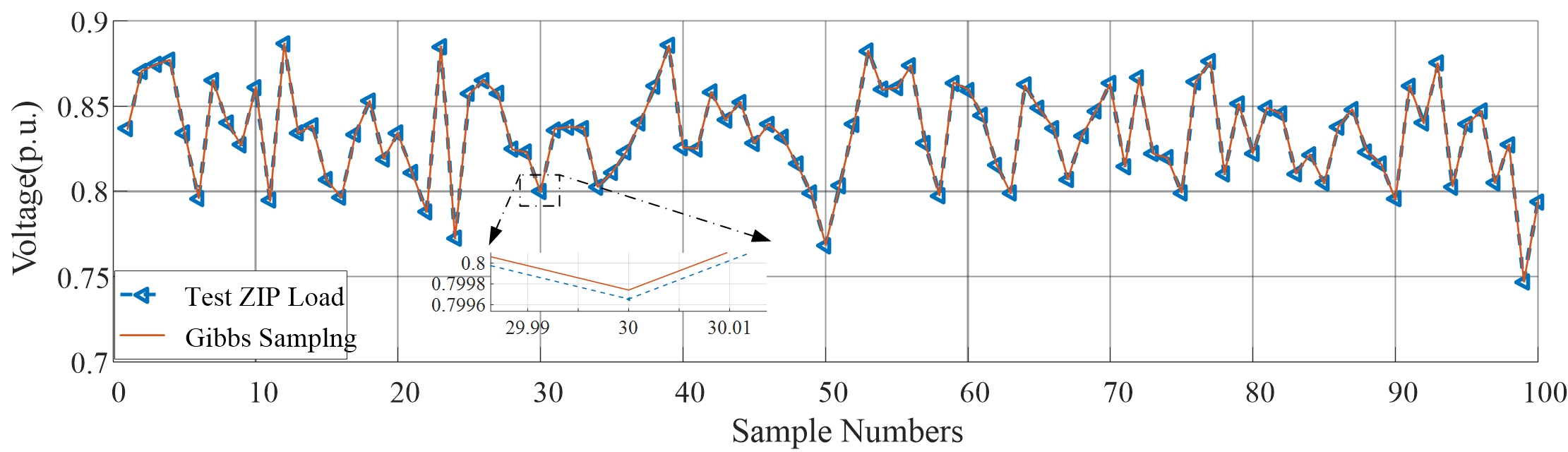}
\caption{Voltage comparison at bus 18 using both the real parameter and estimated parameter.}
\label{comp4}
\end{figure}
\begin{figure}[!t]
\centering
    \includegraphics[width=0.85\linewidth,]{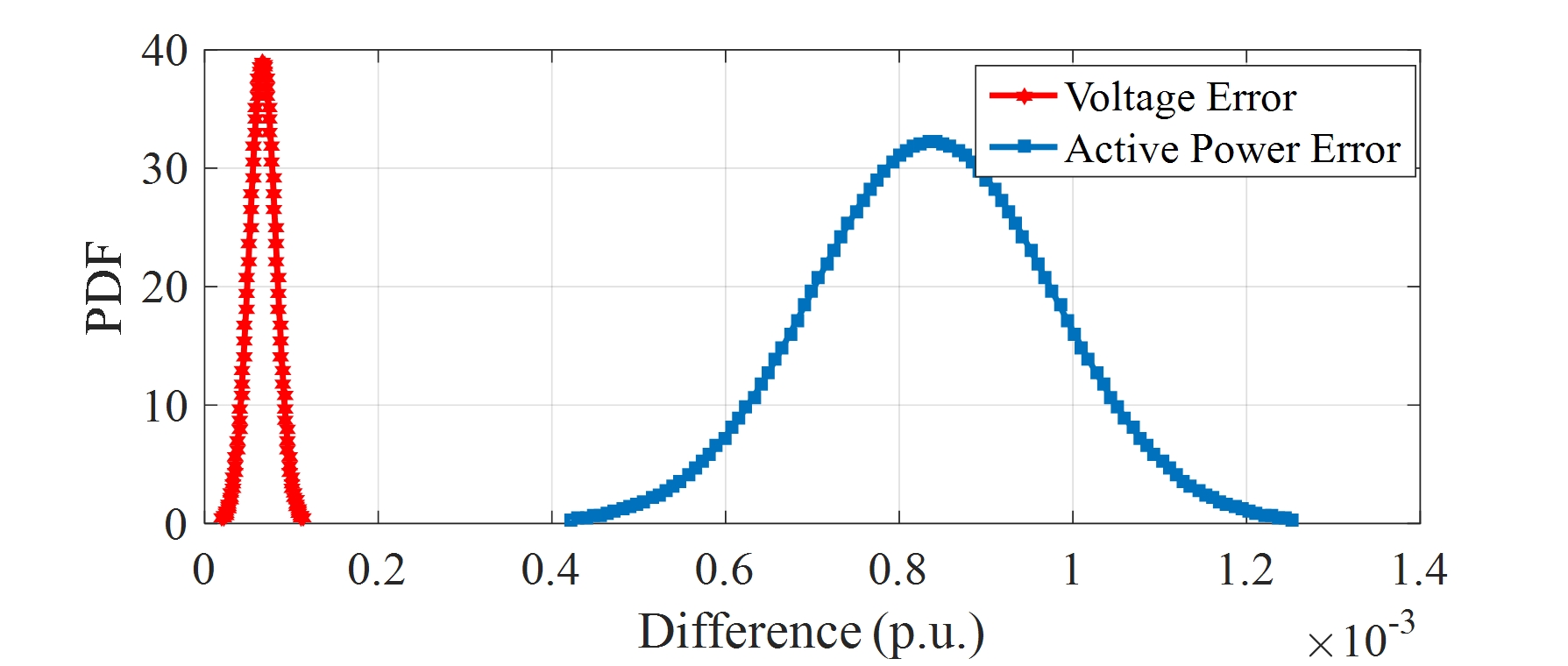}
\caption{Comparison of the absolute values of $\Delta V$ and $\Delta P$ between connecting real ZIP model and the estimated ZIP model at bus 18.}
\label{comp5}
\end{figure}
Different from other estimation approaches, the GS approach can generate a distribution which describes the probability of the real value falling into a certain range. As shown in Fig. \ref{para1}, the mean values of the distributions of $\alpha_1$ and $\alpha_2$ are 0.25 and 0.249, respectively, and $\alpha_3=1-\alpha_1-\alpha_2$. 

\begin{figure}[!t]
\centering
    \includegraphics[width=1\linewidth,]{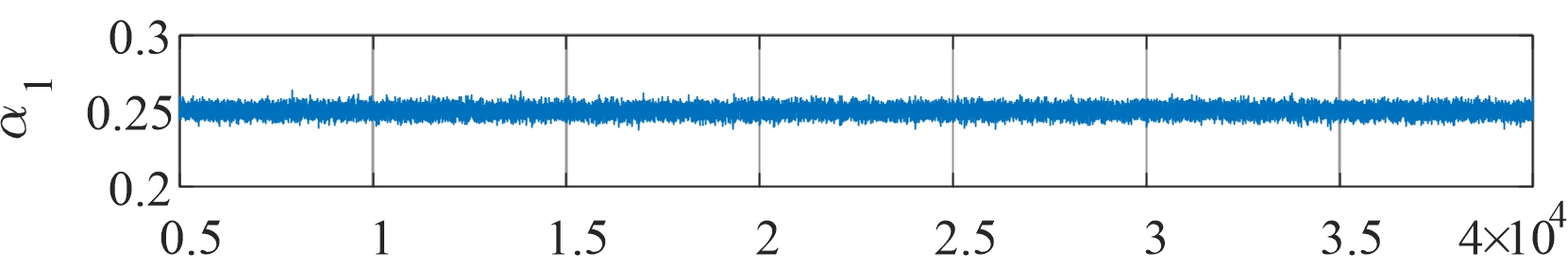}
\caption{The burn-in process of sampling $\alpha_1$.}
\label{burningin}
\end{figure}

     Figs. \ref{comp4} and \ref{comp5} show the voltage and voltage/real power differences, respectively,  using the estimated and real model in 100 i.i.d. experiments. The randomness comes from the factors multiplied to loads at other buses. The dash line with triangle in Fig. \ref{comp4} indicates the voltage at bus 18 using the real ZIP coefficients, and the solid line is the voltage measured at the same bus using estimated coefficients. The comparison shown in Fig. \ref{comp5} is the distributions of voltage and active power differences, $\Delta V$ and $\Delta P$. It can be seen that the differences are in the range of 10$^{-4}$ and 10$^{-3}$, respectively. Fig. \ref{burningin} shows the burn-in process observed. It shows that the process is very fast, and there is no significant burn-in process that can be observed from this figure.
\par

\subsection{IM model Identification}
\begin{figure}[!t]
\centering
    \includegraphics[width=1\linewidth,height=0.29\textheight]{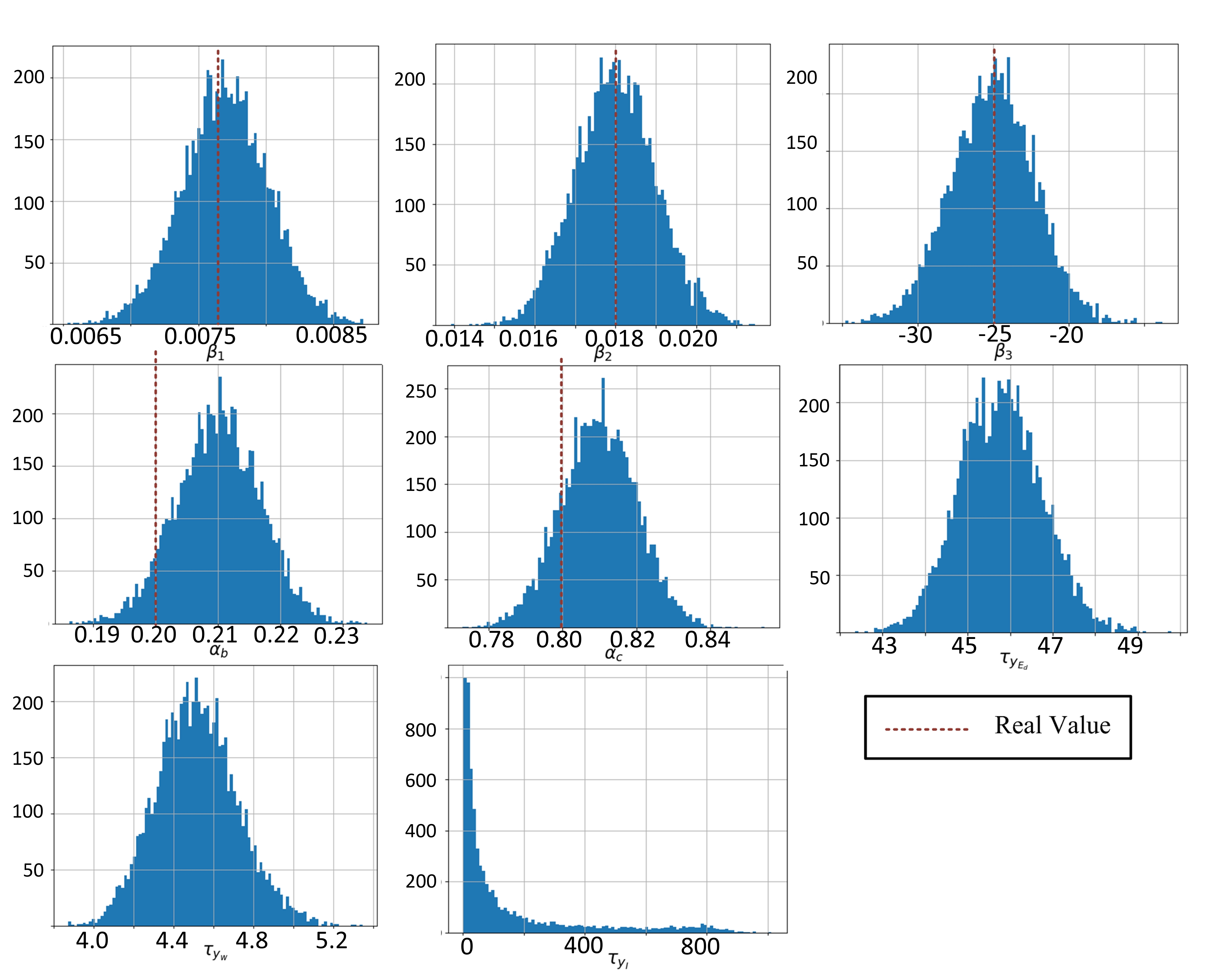}
\caption{The estimated parameter distributions of the IM model.}
\label{imgibbs}
\end{figure}

The estimation of the parameters in the IM model follows the same procedure. A dynamic model is built in Matlab/Simulink to generate the data that is used in (\ref{equa1}). The sampling results are shown in Fig. \ref{imgibbs} and  compared with the real data in Table \ref{imgibbscompare2}.  According to Table \ref{imgibbscompare2}, the estimation error are less than 6\% with 5\% measurements error. It is worth to mention that  estimation of parameters highly relies on the prior distributions. A good prior can significantly increase the estimation accuracy and shorten the burn-in period.
\begin{table}[!t]
\caption{Estimated and real value of the parameters for IM mode with 5\% measurement error}
\label{imgibbscompare2}
\centering
\begin{tabular}{ c|c|c|c }
\hline
Para.&Real Value &Est. Value (mean)&Error(\%)\\
\hline
$\beta_1$&0.0077&0.007683&0.22\\
$\beta_2$&0.018&0.01824&1.33\\
$\beta_3$&25&24.8&0.8\\
$\alpha_b$&0.20&0.211&5.5\\
$\alpha_c$&0.80&0.813&1.63\\
\hline
\end{tabular}
\end{table}\par
\subsection{Benchmarks}
\begin{figure}[!h]
\centering
    \includegraphics[width=1\linewidth]{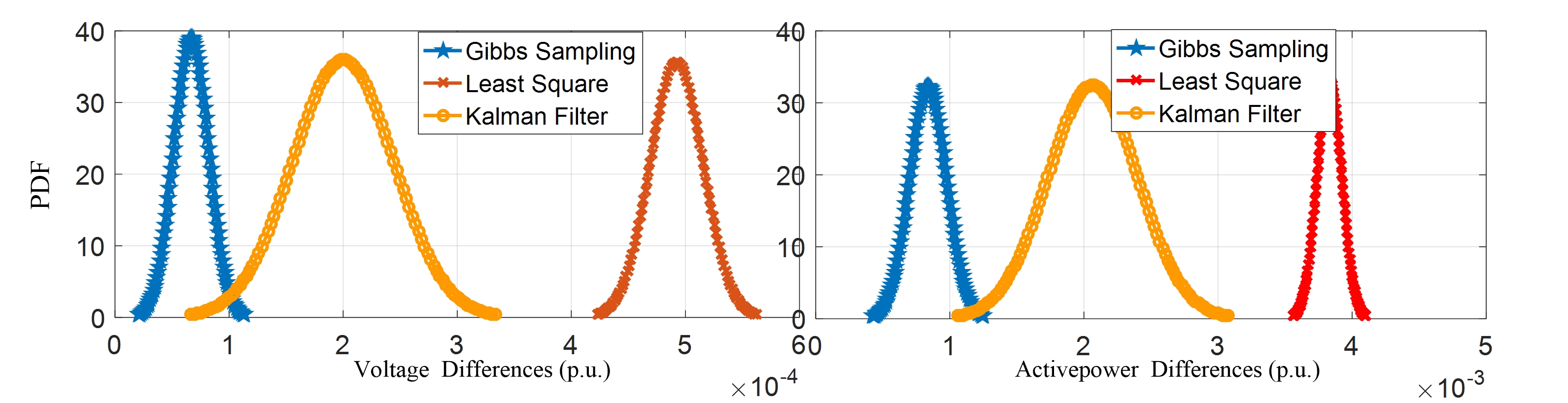}
\caption{The absolute error distributions of voltage and active power using GS, LS and KF, respectively.}
\label{distri}
\end{figure}

The ZIP and IM model parameters derived by the proposed GS method are compared with least square (LS) \cite{27} and Kalman Filter (KF) methods \cite{26} with 10\% noise. The ``fit'' function in Matlab was used to derive the coefficients in LS. The parameter tuning in KF was introduced in \cite{26}, and these parameters were directly used in this study. The comparison results of the voltage and active power using GS, LS, and KF in ZIP model are shown in Fig \ref{distri}. The average absolute mean of voltage and active power errors in p.u. are listed in Table \ref{errors}. According to Fig \ref{distri}, it is obvious that GS approach has the best performance, while KF method only falls slightly behind. In contrast, LS gives the  largest error among all three approaches when there is large measurement error. However, KF only works well with time-invariant parameters estimation \cite{25}.  In practice, the ZIP component varies with time due to stochastic consumer behaviors. \par
\begin{table}[!t]
\caption{Active power \& voltage errors with different methods}
\label{errors}
\centering
\begin{tabular}{ c|c|c|c}
\hline
Para. Error&GS (\%) &LS(\%) &KF(\%) \\
\hline
Voltage&0.007&0.062&0.024\\
Active Power&1.12&4.65&2.64\\
\hline
\end{tabular}
\end{table}\par
The parameters estimation comparisons in the IM model are listed  in Table \ref{imgibbscompare}. The estimation errors from GS are the smallest. The performance of KF is worse than itself in the ZIP model, but still better than the LS method.
\begin{table}[!t]
\caption{IM model parameter estimation with different methods}
\label{imgibbscompare}
\centering
\begin{tabular}{ c|c|c|c|c }
\hline
Para.&GS  &LS&KF&Real Value\\
\hline
$\beta_1$&0.007683&0.0076&0.0077&0.0077\\
$\beta_2$&0.01824&0.1375&0.0185&0.018\\
$\beta_3$&24.8&18&34&25\\
$\alpha_b$&0.211&2.06&1.8&0.2\\
$\alpha_c$&0.813&4.04&3&0.8\\
\hline
\end{tabular}
\end{table}\par
\section{Conclusion}
\label{conclusion}
In this paper, a novel Bayesian estimation based load model parameter identification method is proposed. The proposed method can accurately identify the parameter distributions for both ZIP and IM models. Compared with other measurement based algorithms, the proposed method provides a distribution estimation and is robust to measurement errors. The accuracy and robustness of the proposed method compared with conventional load modeling method are also demonstrated by numerical experiments. The identified distributions can be further used in load prediction, stability and reliability analysis, as well as other related areas. Further study may include how to estimate the ZIP and IM model jointly and extending the process in an online manner.
\bibliographystyle{IEEEtran}
\bibliography{reference}

\end{document}